\documentclass[fleqn]{article}
\begin{document}

{\large
\noindent FUNDAMENTAL INTERACTIONS IN QUANTUM PHASE SPACE\\
SPECIFIED BY EXTRA DIMENSIONAL CONSTANTS 
}

\medskip

{\bf \noindent V.V. Khruschov}\footnote{e-mail: khru@imp.kiae.ru}${}^{,a,b}$

\bigskip

${}^a${\it Russian Research Centre "Kurchatov Institute", Kurchatov Sq. 1,
Moscow 123182}

${}^b${\it Centre for Gravitation and Fundamental Metrology, VNIIMS, 
46 Ozjornaja St., Moscow 119361}

\smallskip

\begin{abstract}
\noindent 
A generalized algebra of quantum  observables,
 depending on extra dimensional constants, is considered.
Some limiting forms of the algebra are investigated and their possible 
applications  to the descriptions of  interactions of
fundamental particles are proposed.  A relation between current
and constituent quark masses is found using a modified
quark equation of Dirac-G\"ursey-Lee type  and  restrictions on the
results of simultaneous measurements of momentum components
are pointed out.
\end{abstract}

\medskip

At present the Standard Model (SM) is the theory of three fundamental interactions,
namely, strong, electromagnetic and weak ones. The theory is based on
 definitions of quantum fields in the continuous Minkowski space-time (MS).
The group of MS motions is the Poincar{\'{e}} group, its irreducible
unitary representations (IRs) are used to specify elementary particles.
This procedure  works successfully for all visible particles
such as protons and electrons, but it is not well justified for unusual
particles (UPs) such as quarks. UPs can exist in states of
matter under extreme conditions, for instance, in the early Universe, in the
quark-gluon plasma or inside of  elementary particles \cite{Coles, Silk}. 
Space-time properties of UPs may be
more complicated and  determined by a generalized group of
space-time symmetries in extra dimensions \cite{Akama, Rub}. In this case an  algebra of  observables of
quantum theory  can depend on additional fundamental constants as
compared with the light velocity $c$ and the Planck constant of action $\hbar$.

In this paper we consider  a generalized  algebra of  observables for
quantum theory, which depends on extra fundamental constants 
with dimensions of mass, length and action  \cite{Lez-Kh}. It is known 
the conventional quantum theory has a long-standing problem concerning the removal of
short-distance singularities. It was the reason which inspired Heisenberg to
suggest the idea that the configuration-space coordinates may not commute
\cite{Jack}. Then Snyder introduced a Lorentz-invariant quantized space-time
characterized by a fundamental length \cite{Sny}.  This theory is, however, not
invariant under translations.  A generalized translation 
invariance, which leads to noncommutative
momenta and so to a new fundamental unit of mass,  was suggested by Yang 
\cite{Ya}.  The most general algebra of  quantum observables 
  has been found providing the  Lorentz invariance in Ref.\cite{Lez-Kh} 
(see also Ref.\cite{Kh-Lez}) 
and a new fundamental constant with the dimensions of action has been 
introduced. 

Let us write  generalized commutation relations in the form presented in 
Refs.\cite{Lez-Kh, Kh-Lez} for 
 operators $F_{ij}$, $p_{i}$, $x_{i}$ and $I$,
 $x_{i}$ and $p_{i}$ are  operators of  4-coordinates and 4-momenta,
respectively, $F_{ij}$ are  proper Lorentz group generators, $I$ is a
so-called "unit operator". 
\[
\hspace*{1.8cm}[F_{ij}, F_{kl}]=if(g_{jk}F_{il}-g_{ik}F_{jl}+g_{il}F_{jk}-g_{jl}F_{ik}),
\]
\[
\hspace*{1.8cm}[p_{i}, x_{j}]=if(g_{ij}I+\frac{F_{ij}}{H}),
\]
\[
\hspace*{1.8cm}[p_{i}, p_{j}]=\frac{if}{L^{2}}F_{ij},
\]
\[
\hspace*{1.8cm}[x_{i}, x_{j}]=\frac{if}{M^{2}}F_{ij},
\]
\[
\hspace*{1.8cm}[p_{i}, I]  =  if\biggl(\frac{x_{i}}{L^{2}}-\frac{p_{i}}{H}\biggr),
\]
\[
\hspace*{1.8cm}[x_{i}, I]  =  if\biggl(\frac{x_{i}}{H}-\frac{p_{i}}{M^{2}}\biggr),
\]
\[
\hspace*{1.8cm}[F_{ij}, p_{k}]  =  if(g_{jk}p_{i}-g_{ik}p_{j}),
\]
\[
\hspace*{1.8cm}[F_{ij}, x_{k}]  =  if(g_{jk}x_{i}-g_{ik}x_{j}),
\]
\begin{equation}
\hspace*{1.8cm}[F_{ij}, I]  =  0.
\label{finalg}
\end{equation}

The algebra (\ref{finalg}) depends on four dimensional parameters: $L$ is a
constant with the dimensions of length, $M$ with the dimensions of mass, $H$
and $f$ with the dimensions of action ($M$ and $L$ can take real values as well
as pure imaginary ones, $c=1$ in the system of units being used).
In the general case, the algebra (\ref{finalg}) can be considered
as the algebra of observables for some Lorentz-invariant quantum theory with
noncommutative coordinates and momenta.  The commutation relations (\ref{finalg}) go
over into the commutation relations of the canonical (at present)
quantum field theory providing
$f=\hbar $ in the limiting case, when $M$, $L$
and $H$ become infinitely large.  Note that a more complicated case is also possible, 
when $f$ is some function $f(L,M,H)$,  which tends to $\hbar$ in the 
limiting case, as $L\to \infty $, $M\to \infty $ and $H\to \infty $.

The system of commutation relations
(\ref{finalg}) specifies some class of Lie algebras which consists of
semisimple algebras as well as general-type algebras. On performing
the calculation of the Killing-Cartan form the condition of 
semisimplicity can be written as
\begin{equation}
\hspace*{3.4cm} \frac{f^{2}(M^{2}L^{2}-H^{2})}{H^{2}M^{2}L^{2}}\neq 0.    
\label{killing}
\end{equation}
If one carries out a linear transformation of the $p_{i},\ x_{i},\ I$
generators, which has the form
\[
\hspace*{3.4cm}F_{i5} =  Bx_{i}+Dp_{i},
\]
\begin{equation}
\hspace*{3.4cm}	F_{i6} = Ex_{i}+Gp_{i},\quad F_{56}=AI,
\end{equation}
then one can obtain the commutation
relations for the algebras of pseudo-orthogonal groups $O(3,3)$, $O(2,4)$ and
$O(1,5)$  under the condition (\ref{killing}).  These algebras correspond to specific values of the parameters
$M^{2}$, $L^{2}$ and $H^{2}$ \cite{Lez-Kh, Kh-Lez}.

IRs of the algebras (\ref{finalg}) are determined
with the help of eigenvalues of Casimir operators. For the  pseudoorthogonal 
groups in six-dimensional spaces  the Casimir operators have the known forms in
terms of the generators $F_{ij}$, $i,j=0,1,...,5 $:
\[
\hspace*{3.4cm}	K_1 = \epsilon_{ijklmn}F^{ij}F^{kl}F^{mn},
\]
\[
\hspace*{3.4cm} K_2 = F_{ij}F^{ij},
\]
\begin{equation}
\hspace*{3.4cm} K_3 = (\epsilon_{ijklmn}F^{kl}F^{mn})^2.
\label{cas}
\end{equation}
$K_1$, $K_2$ and $K_3$ can also be expressed in terms of the $I$,
$p_i$, $x_i$, $F_{ij}$, $i,j=0,...,3$,   operators. For instance, the second-order
invariant operator $K_2$, which in terms of  
the $I$, $p_i$, $x_i$, $F_{ij}$  we denote as $C_2$,
can be presented as
\begin{equation}
\hspace*{1.8cm}	 C_2=\sum_{i<j}F_{ij}F^{ij}(\frac{1}{M^{2}L^{2}}-
			\frac{1}{H^{2}})+I^{2}
	+\frac{x_{i}p^{i}+p_{i}x^{i}}{H}-
		\frac{x_{i}x^{i}}{L^{2}}-\frac{p_{i}p^{i}}{M^{2}}.
\label{casim}
\end{equation}
$C_2$ in the limiting case $M\to \infty $, $L\to \infty ,$ $H\to \infty $
transforms into the  "unit operator" $I$ squared.

One can apply the algebra (\ref{finalg}) for a description of  
megaphenomena. In this case values of the parameters entering into the 
system (\ref{finalg}) should be of cosmic scales under the condition that
inviolate physical phenomena take place at least at distances of the order of 
the Solar system. Possible applications of the algebra (\ref{finalg})
for such phenomena have  been developed in the Refs.\cite{Lez1, Lez2}. In 
the Refs.\cite{Kh1, Kh2} the algebra (\ref{finalg}) has been applied for microphenomena. 
In what follows we assume that the algebra (\ref{finalg}) is suitable for a description 
of UPs such as quarks or preons.

Let us define new microscopical constants $\kappa$, $\lambda$ and $\mu$ 
by means of constants $M$, $L$, $H$ and $f$.
First of all we set $f = \hbar$ because of   the algebra (\ref{finalg}) 
is applied for microphenomena. Then it is convenient to define
\begin{equation}
\hspace*{3.4cm} \kappa=\hbar/H, \quad \lambda = \hbar/M, \quad \mu=\hbar/L.
\end{equation}
So  the algebra (\ref{finalg}) can be written in
 the natural units with $c=\hbar=1$ as
\[
\hspace*{1.8cm}	[F_{ij},F_{kl}] =
   		i(g_{jk}F_{il}-g_{ik}F_{jl}+g_{il}F_{jk}-g_{jl}F_{ik}),
\]
\[
\hspace*{1.8cm}	[p_{i},x_{j}] = i(g_{ij}I+\kappa{F_{ij}}),
\]
\[
\hspace*{1.8cm}	[p_{i},p_{j}] = {i}{\mu^{2}}F_{ij},
\]
\[
\hspace*{1.8cm}	[x_{i},x_{j}] = {i}{\lambda^{2}}F_{ij},
\]
\[
\hspace*{1.8cm}	[p_{i},I] = i(\mu^{2}{x_{i}}-\kappa{p_{i}}),
\]
\[
\hspace*{1.8cm}	[x_{i},I] = i(\kappa{x_{i}}-\lambda^{2}{p_{i}}),
\]
\[
\hspace*{1.8cm}	[F_{ij},p_{k}] = i(g_{jk}p_{i}-g_{ik}p_{j}),
\]
\[
\hspace*{1.8cm}	[F_{ij},x_{k}] = i(g_{jk}x_{i}-g_{ik}x_{j}), 
\]
\begin{equation}
\hspace*{1.8cm}	[F_{ij},I] = 0.
\label{fina}
\end{equation}

Some physical interpretations and  mathematical 
properties of an algebra, which corresponds to the algebra (\ref{fina}),  
 are considered  in the Ref.\cite{Ch-Ok}. 
We shall restrict our consideration to the application of the algebra (\ref{fina}) 
to account for quarks. It is known the presence of the nonzero $\kappa$ leads
to the $CP-$violation \cite{Lez-Kh, Kh-Lez}, because of this 
it has been noted  \cite{Kh2}, that the 
condition $\kappa = 0$ should be hold for quarks due to the fact 
 the strong interactions   are invariant 
with respect to the $P-$, $C-$ and $T-$transformations on the high level 
of precision.  Moreover, the presence of the nonzero $\lambda$
value  causes to some inconsistencies for the description 
of quarks and is superfluous. It can be easily seen, if one take into 
account that the constant $\lambda$ should be most likely connected with a typical size 
of the confinement domain  for quarks and gluons, which is of the order of 
hadron size. Thus, we put $\kappa=$$\lambda = 0$ and denote $\mu$ as $\mu_s$. 
In this case the particular form of the  algebra (\ref{fina}) can be written for strong
interacting fundamental particles such as quarks and gluons in the 
 form presented below,  the unaltered relations from  algebra (\ref{fina})
(for $[F_{ij},F_{kl}]$, $[F_{ij},p_{k}]$, $[F_{ij},x_{k}]$, $[F_{ij},I]$) are
not shown.
\[
\hspace*{1.8cm}	[p_{i},x_{j}] = ig_{ij}I,
\]
\[
\hspace*{1.8cm}	[p_{i},p_{j}] = {i}{\mu_s^{2}}F_{ij},
\]
\[
\hspace*{1.8cm}	[p_{i},I] = i\mu_s^{2}{x_{i}},
\]
\[
\hspace*{1.8cm}	[x_{i},x_{j}] = 0,
\]
\begin{equation}
\hspace*{1.8cm}	[x_{i},I] = 0.
\label{fins}
\end{equation}

The direct consequence  the commutation relations written above is the 
rise of nonzero standard uncertanties for quark momentum components
in simultaneous measurements. For instance, let 
$\psi_{1/2}$ be a quark state with the definite value of its spin component
along the third axis. Then   $[p_1, p_2] = i\hbar \mu_s/2$, as it follows 
from commutation relation for $p_1$  and $p_2$, so  
\begin{equation}
\hspace*{3.4cm}	\Delta p_1  \Delta p_2 \ge \mu_s^2/4,
\label{ineq}
\end{equation}
\noindent and if $\Delta p_1 \sim   \Delta p_2$ one gets
$ \Delta p_1\ge \mu_s/2$, $\Delta p_2 \ge \mu_s/2$. Taking into account 
the  value of $\mu_s$ estimated  further, we see
the generalized quark components cannot be measured better than tentatively
one-half a mass of the $\pi-$meson.

Rough estimations in the framework of quark model give us the $\mu_s$ value
should be in the neighborhood of $0.5$ GeV. For instance, we can use a quark
equation  of the Dirac-G\"ursey-Lee type \cite{Dir, Gu-Le}
\begin{equation}
\hspace*{2.2cm} [\gamma_i(p_0^i + dp_0^kL_k^i + i\mu_s\gamma^i/2) + 
2i\mu_sS_{ij}(L^{ij}+S^{ij})]\psi = m\psi,
\label{equ}
\end{equation}

\noindent   the $p^i$ operator is equal to $p_F^i +i\mu_s\gamma^i/2$
and following to the Ref.\cite{Fron} $p_F$ is a space-time part 
of the total momentum: $p_F^i = p_0^i + dp_0^kL_k^i$, $d = \mu_s/m_0$
and $p_0$, $L_{ij}$ are the usual generators of translations and Lorentz 
transformations in Minkovski space-time, $p_0^2 = m_0^2$. We assume that 
the mass value of a constituent quark arise predominantly due to the noncommutativity
of $p_i$ compoments as is shown in the relations (\ref{fins}). So we obtain 
for the quark state $\psi$ with $ L^{ij}\psi$
 equal to zero the relation, which provides a possibility
to estimate the magnitude of $\mu_s$:
\begin{equation}
\hspace*{4.2cm}	m \approx m_0 + 2i\mu_s,
\label{const}
\end{equation}
\noindent $m_0$ can be identify as the current quark mass, while $m$ as
the constituent quark mass. For example one can use for the 
current and constituent masses of
 u-quark the values 2 MeV and 316 MeV, respectively, whereas the 
energy of the constituent quark is equal to 335 MeV in a hadron ground state 
\cite{Yao, Ko-Me-Kh, Kh}. Thus $\mu_s$  
should be pure imaginary negative and $|\mu_s| \approx 
157$ MeV. This value is one-half the value for $\mu_s$, which has been 
obtained by other means in Ref.\cite{Kh2}. But two constants $\mu_s$ and $\lambda_s$  
have been used in this work, what is superfluous for strong interactions 
of quarks and gluons, as it is pointed out previously.  
The algebra (\ref{fins}) is isomorphic to the 
algebra of the AdS group O(2,3) due to the condition $\mu_s^2<0$.

As it follows immediately from Eq.(\ref{const}) the current and constituent
masses for all quarks differ approximately from one another  by the
constant term, which is equal to  $2|\mu_s|$.  
Taking into account the values of constituent masses presented in 
Refs.\cite{Ko-Me-Kh, Kh}, we can deternimate
the current  masses for $ d-$, $s-$, $c-$ and $b-$quarks 
 on the scale of their constituent masses:
$m_{cur}^d(m_{con}^d)\approx 6.2 $ MeV,
$m_{cur}^s(m_{con}^s)\approx 158 $ MeV,
$m_{cur}^c(m_{con}^c)\approx 1292 $ MeV,
$m_{cur}^b(m_{con}^b)\approx 4635 $ MeV.

In  summary let us make one additional comment.
Very likely the Universe has a quantum origin at the beginning.
So the dispersion relation between an energy and a momentum
for a cosmic system can be depended on its characteristics
 on a quantum stage of formation, with regard for the numerical 
values of constants can variate during a certain formation period
(see, e.g. \cite{Me, Uz}). This being so  the relation (\ref{const}) 
can be realized and  effective masses of cosmic 
systems can  increase analogously to increasing of quark masses.

\smallskip

{\bf \noindent Acknowledgments.} The author is grateful A.N. Leznov,
S.A. Kononogov, Yu.V. Gaponov, V.N. Melnikov, M.I. Kalinin, S.V. Semenov and
A.M. Snigirev for useful and stimulating discussions. The work 
was partially supported by RFBR grant 07-02-13614-ofi-ts.


\small
\begin{thebibliography}{99}

\bibitem {Coles}  P. Coles and F. Lucchin, "Cosmology - 
The Origin and Evolution of  Cosmic Structure",
West Sussex:John Wiley and Sons Ltd, 1995.

\bibitem {Silk}  J. Silk, ``The Big Bang'',  3rd ed., W.H. Freeman \& Co.,
 2001.

\bibitem {Akama}  K. Akama,
 		{\it Lect. Notes Phys. } {\bf 176}, 267 (1983).

\bibitem {Rub}  V.A. Rubakov and  M.E. Shaposhnikov,
 		{\it Phys. Lett.} {\bf 125B}, 136 (1983).

\bibitem {Lez-Kh}  A.N. Leznov and  V.V. Khruschev,
 		{\it preprint } IHEP 73-38, Serpukhov, 1973.

\bibitem {Jack}  R. Jackiw, {\it Lochlainn O'Raifeartaigh, Fluids,
and Noncommuting Fields}, arXiv:physics/0209108.

\bibitem {Sny}  H. Snyder, {\it Phys. Rev. } {\bf 71}, 38  (1947).

\bibitem {Ya}  C.N. Yang, {\it Phys. Rev. } {\bf 72}, 874 (1947).

\bibitem {Kh-Lez}   V.V. Khruschov and A.N. Leznov, {\it Grav. Cosmol.}
{\bf 9}, 159 (2003) [arXiv:hep-th/0207082].

\bibitem {Lez1}  A.N. Leznov, {\it Free motion in deformed
(quantum) four-dimensional space}, arXiv:0707.3420.

\bibitem {Lez2}  A.N. Leznov, {\it Kepler problem in deformed
(quantum) four-dimensional space in non relativistic limit
with Galilei group of motion}, arXiv:0708.0643.

\bibitem {Kh1}  V.V. Khruschov, {\it Measurement Techniques}
{\bf 37}, no 7, 717 (1994).

\bibitem {Kh2}  V.V. Khruschov, {\it Grav. Cosmol.} {\bf 3}, 197 (1997).

\bibitem {Ch-Ok}  C. Chryssomalakos, E. Okon, {\it Int.J.Mod.Phys.}
{\bf D13}, 2003 (2004) [arXiv:hep-th/0410212].

\bibitem {Dir}  P.A.M. Dirac, {\it Ann. Math. } {\bf 36}, 657
	(1935).

\bibitem {Gu-Le}  F. G\"ursey and T.D. Lee, {\it Proc. Nat. Sc. Sci.}
	{\bf 49}, 179 (1963).

\bibitem {Fron}  C. Fronsdal,  {\it Rev. Mod. Phys.} {\bf 37}, 221 (1965).

\bibitem {Yao}  W.-M. Yao et al.(Particle Data Group),  
{\it J. Phys. G} {\bf 33}, 1 (2006).

\bibitem {Ko-Me-Kh} S.A. Kononogov, V.N. Melnikov, V.V. Khruschov, 
{\it Izmeritelnaya tekhnika,} {\bf3}, 3 (2007).

\bibitem {Kh}  V.V. Khruschov, The lecture at the 
XLI PNPI Winter School, 2007, Repino, arXiv: hep-ph/0702259.

\bibitem {Me}  V.N. Melnikov, {\it Grav. Cosmol.} {\bf 13}, 81 (2007).

\bibitem {Uz}  J.-P. Uzan,  {\it Rev. Mod. Phys.} {\bf75}, 403 (2003).

\end{thebibliography}
\end{document}